\documentstyle[12pt]{article}
\begin{document}
\mbox{ }
\rightline{UCT-TP-232/96}
\rightline{June 1996}
\vspace{3.0cm}
\begin{center}
\begin{Large}
{\bf  Pions at Finite Temperature from QCD Sum Rules} 
\end{Large}

\vspace{1cm}

{\large {\bf C. A. Dominguez}$^{a}$},
{\large {\bf M. S. Fetea}$^{a}$}, and
{\large {\bf M. Loewe}$^{b}$}
\end{center}

\vspace{.5cm}

\begin{center}
\begin{tabular}{lc}
$a$&Institute of Theoretical Physics and Astrophysics, University of Cape Town,\\
    &Rondebosch 7700, South Africa\\[.3cm]
$b$&Facultad de Fisica, Pontificia Universidad Catolica de Chile,\\
    &Casilla 306, Santiago 22, Chile
\end{tabular}
\end{center}
\vspace{1cm}

\begin{abstract}
\noindent
The temperature corrections to the current algebra Gell-Mann, Oakes, and
Renner (GMOR) relation in $SU(2)\otimes SU(2)$
are investigated in the framework of QCD sum rules.
There are no corrections at leading order in the quark masses. At the next
to leading order we find corrections of the form $m_{q}^2\; T^{2}$, which
are small except near the critical temperature. As a by-product we obtain
the temperature behaviour of the pion mass, which is essentially constant,
except near the critical temperature where it increases with $T$.
\end{abstract}
\newpage
\setlength{\baselineskip}{1.5\baselineskip}
\noindent 
Due to its small mass, the pion plays a special role in the dynamics
of hot hadronic matter. Therefore, it is quite important to understand
the temperature behaviour of the pion's Green function. The pion mass
$\mu_{\pi}(T)$ has been studied in a variety of frameworks, such as
Chiral Perturbation Theory (low temperature expansion) \cite{GL},
the Linear Sigma Model \cite{LCL}, the Mean Field Approximation \cite{Bar1},
the Virial Expansion \cite{Schenk}, etc. There seems to be a reasonable
consensus that $\mu_{\pi}(T)$ is essentially independent of $T$, except 
possibly near the critical temperature $T_{c}$ where $\mu_{\pi}(T)$ increases 
with $T$. While the pion is (hadronically) stable at $T = 0$, it is expected
to develop a width (imaginary part of the Green function) at 
non-zero temperature, such width being interpreted as a damping coefficient
which should diverge at the critical temperature for deconfinement.
This follows from a proposal   \cite{CAD1} - \cite{CAD2} (see
also \cite{PIS}) to consider the width of a hadron as a phenomenological 
order parameter for the deconfinement phase transition. In fact, as the
temperature is increased and the hadron melts, its width should increase
until it becomes infinite at $T=T_{c}$, thus ensuring that
no resonance peaks remain in the hadronic spectral function. The latter
should become a smooth function of the energy and coincide with its
perturbative QCD value. These properties
have been confirmed by detailed calculations in
the framework of the virial expansion \cite{LS} and the Linear Sigma
Model \cite{CAD3}. \\

In this paper we study the temperature corrections to the well known
Gell-Mann, Oakes, and Renner (GMOR) relation of current algebra
\cite{GMOR}
\begin{equation}
f_{\pi}^{2} \; \mu_{\pi}^{2} = - (m_{u} + m_{d}) \; <\bar{q}q>
\end{equation}
where $f_{\pi} \simeq$ 93 MeV, and $<\bar{u} u> \simeq <\bar{d} d>
\equiv <\bar{q} q> \simeq$ -- 0.01 GeV$^{3}$. We do this in the framework of
QCD sum rules at finite temperature \cite{CAD1}, \cite{BS}, \cite{CAD4}.
As a byproduct of our analysis, we obtain the temperature dependence of
the pion mass.\\

We begin by considering the following two-point function (at $T = 0$)
\begin{equation}
\Pi_{5 \mu} (q) = i \int \; d^{4}x \; e^{iq \cdot x} < 0| \; 
T \; (A_{\mu} (x) \; j_{5}^{\dag} \; (0)) \; |0> \; = \; iq_{\mu} \; 
\Pi_{5} (q^{2}) \; ,\end{equation}
where $A_{\mu}(x) = : \bar{d}(x) \; \gamma_{\mu} \; \gamma_{5} \; 
u(x): \; $, and $j_{5} (x) = :\bar{d}(x) \; i  \gamma_{5} \; u(x):$. 
Calculating $\Pi_{5}(q^{2})$ in perturbative QCD to order
${\cal{O}}$$(m_{q}^{3})$, and incorporating the leading non-perturbative 
corrections, we find
\begin{eqnarray}
\Pi_{5} \; (q^{2})|_{\mbox{QCD}} \; &=& - \; \frac{3}{8 \pi^{2}}
\; (m_{u} + m_{d}) \; \ln(-q^{2}/\nu^{2}) + \frac{3}{4 \pi^{2}} \; m_{u} \; m_{d} \;  
(m_{u} + m_{d}) \; \frac{1}{q^{2}}\nonumber\\[.3cm]
 & & + \; \frac{2<\bar{q}q>}{q^{2}} + \frac{1}{2} \; (m_{u} + m_{d})^{2} \;
     \frac{<\bar{q}q>}{q^{4}} - \frac{1}{8} \;
     \frac{(m_{u} + m_{d})}{q^{4}} <\frac{\alpha_{s}}{\pi} G^{2}> \; 
     .\nonumber\\
 & & 
\end{eqnarray}
Using $m_{u} \simeq$ 5 MeV, $m_{d} \simeq$ 10 MeV  \cite{QM}, and
$< \frac{\alpha_{s}}{\pi} \; G^{2} > \simeq$ (1 - 4) $\times$ 10$^{-2}$ 
GeV$^{4}$
\cite{GC}, the above expression can be safely approximated for
$- q^{2} \simeq$ 1 GeV$^{2}$ as
\begin{eqnarray}
\Pi_{5} \; (q^{2})|_{\mbox{QCD}} \; &\simeq& \frac{-3}{8 \pi^{2}}
\; (m_{u} + m_{d}) \; \ln \; (-q^{2}/\nu^{2})\nonumber\\[.3cm] 
& & + \; \frac{2<\bar{q}q>}{q^{2}} - \frac{1}{8} \;
     \frac{(m_{u} + m_{d})}{q^{4}} \; < \frac{\alpha_{s}}{\pi} \; 
     G^{2} > \; .
\end{eqnarray}
On the hadronic side, saturation with the lowest hadronic state (the 
pion) yields
\begin{equation}
\Pi_{5} \; (q^{2})|_{\mbox{HAD}} = 
\frac{2 f_{\pi}^{2} \; \mu_{\pi}^{2}}{(m_{u} + m_{d})} \;
\frac{1}{\mu_{\pi}^{2} - q^{2}} \; .
\end{equation}
We shall ignore in the sequel contributions from higher resonances
($\pi', \pi''$, etc.) because the analysis will be basically 
restricted to energies below 1 GeV, in which case these hadronic 
contributions are absorbed into the continuum.

Invoking Cauchy's theorem in the form
\begin{equation}
\frac{1}{\pi} \int_{0}^{s_0} \; s^{N} \; \mbox{Im} \; 
\Pi_{5} (s)|_{\mbox{HAD}} \; ds \; = \; - \frac{1}{2 \pi i} \;
\int_{C(|s_{0}|)} \; s^{N} \; \Pi_{5} (s)|_{\mbox{QCD}} \; ds,
\end{equation}
where $N$ = 0,1,2 $\cdots$, leads to Finite Energy Sum Rules (FESR). 
The first two FESR in our case read
\begin{equation}
2 f_{\pi}^{2} \; \mu_{\pi}^{2} \; = \; - 2 (m_{u} + m_{d}) \;
<\bar{q}q> + \frac{3}{8 \pi^{2}} \; (m_{u} + m_{d})^{2} \;
\int_{0}^{s_{0}} \; ds,
\end{equation}
\begin{equation}
2 f_{\pi}^{2} \; \mu_{\pi}^{4} \; = \; \frac{1}{8} (m_{u} + m_{d})^{2} \;
<\frac{\alpha_{s}}{\pi} \; G^{2}> + \frac{3}{8 \pi^{2}} \; (m_{u} + m_{d})^{2} \;
\int_{0}^{s_{0}} \; s \; ds,
\end{equation}
where $s_{0}$ is the continuum threshold. Equation (7) becomes the 
GMOR relation at leading order in the quark masses. We have left
the explicit form of the trivial integrals above, in order to compare
later with the FESR at finite temperature, where these integrals cannot
be calculated analytically in closed form.

Next, we reconsider the above FESR at finite temperature. Thermal 
corrections to $\Pi_{5} \; (q^{2})|_{\mbox{QCD}}$ can be calculated 
in the standard fashion \cite{CAD1}, \cite{BS},  \cite{CAD4}, and we 
find for the imaginary part
\begin{eqnarray}
\mbox{Im} \; \Pi_{5} \; (s,T)|_{\mbox{QCD}} &=& \frac{3}{8 \pi}
(m_{u} + m_{d}) \; \left[ 1 - 2 \; n_{F} \left( \frac{\sqrt{s}}{2T} 
\right) \right]\nonumber\\[.3cm]
 & & + \;  \frac{\pi}{2} \; (m_{u} + m_{d}) \; T^{2} \; \delta(s), 
\end{eqnarray}
where $n_{F}(x) = (1 + e^{x})^{-1}$ is the Fermi thermal factor. In 
addition, the non-perturbative vacuum condensates will develop a 
temperature dependence. For $<\bar{q}q>_{T}$ we shall use the results 
of \cite{Bar1} away from the chiral limit, i.e. for $m_{q} \neq 0$. The 
gluon condensate is basically independent of $T$, except very close 
to the critical temperature $T_{c}$ \cite{GEL}, so that we shall take 
it as a constant. On the hadronic side, the pion mass and decay constant
will develop a 
temperature dependence, and so will $s_{0}$. The latter follows from the
notion that as the resonance peaks in the spectral function become broader,
the onset of the continuum should shift towards threshold
\cite{CAD1}, \cite{Bar2}. The temperature behaviour of this asymptotic
freedom threshold can be obtained from the lowest dimension FESR associated 
to the two-point function involving the axial-vector currents 
\cite{CAD1}, \cite{Bar2}, provided $f_{\pi} (T)$ is known, viz.
\begin{eqnarray}
8 \pi^{2} \; f_{\pi}^{2} (T) &=& \frac{1}{2} \; \int_{0}^{s_{0}(T)} \;
d z^{2} \; v(z) \; [3 - v^{2}(z)] \; \left[ 1 - 2n_{F} \left( \frac{z}{2 T}
\right) \right]\nonumber\\[.3cm]
& & + \; \int_{0}^{\infty} \; d z^{2} \; v(z) \; [3 - v^{2}(z)]
 \; n_{F} \left( \frac{z}{2T} \right),
\end{eqnarray}
where $v(z) = (1 - (m_{u} + m_{d})^{2}/z^{2})^{\frac{1}{2}}$.
We have used $f_{\pi}(T)$ as determined in \cite{Bar1} (for $m_{q} 
\neq 0$), and solved the above FESR for $s_{0}(T)$. The result is 
shown in Fig.1, together with the input $f_{\pi}(T)$, as well as
$<\bar{q}q>_{T}$, both from \cite{Bar1}. It is interesting to notice that
for temperatures not too close to $T_{c}$, say $T < 0.8 T_{c}$, the
following scaling relation holds to a good approximation
\begin{equation}
\frac{f_{\pi}^{2}(T)}{f_{\pi}^{2}(0)} \simeq
\frac{<\bar{q}q>_{T}}{<\bar{q}q>} \simeq
\frac{s_{0}(T)}{s_{0}(0)}.
\end{equation}
The FESR, Eqs.(7) and (8) now become
\begin{eqnarray}
G(T) &\equiv& 2 f_{\pi}^{2}(T) \; \mu_{\pi}^{2}(T) + 2 \; (m_{u} + 
m_{d}) \; <\bar{q}q>_{T}\nonumber\\[.3cm]
&=& \frac{3}{8 \pi^{2}} \; (m_{u} + m_{d})^{2} \; \left\{
\frac{4 \pi^{2}}{3} \; T^{2} + \int_{0}^{s_{0}(T)} \; ds \;
\left[ 1 - 2 n_{F} \left( \frac{\sqrt{s}}{2T} \right) \right] 
\right\} \; ,
\end{eqnarray}

\begin{equation}
2 f_{\pi}^{2}(T) \; \mu_{\pi}^{4}(T) = \frac{3}{8 \pi^{2}} \; (m_{u} + m_{d})^{2} \; \left\{
\frac{\pi}{3} <\alpha_{s} G^{2}> + \int_{0}^{s_{0}(T)} \;\; 
\left[ 1 - 2 n_{F} \left( \frac{\sqrt{s}}{2T} \right) \right] 
s \; ds \right\} \; .
\end{equation}

In Fig.2 we show the ratio
\begin{equation}
\frac{G(T)}{G(0)} = \frac{4 \pi^{2}}{3} \; \frac{T^{2}}{s_{0}(0)} \;
+ \frac{1}{s_{0}(0)} \;\int_{0}^{s_{0}(T)} \; ds \;
\left[ 1 - 2 n_{F} \left( \frac{\sqrt{s}}{2T} \right) \right] 
\end{equation}
for $s_{0}(0)$ = 1 GeV$^{2}$ (reasonable changes around this value 
have basically no influence on the results). Qualitatively, our result is
in agreement with expectations: since both $f_{\pi}(T)$ and $<\bar{q}q>_{T}$
decrease with $T$, the same should be true of $G(T)$.
From the point of view of the sum rule, Eq.(14),
the decrease of $G(T)$ is due to the decrease of $s_{0}(T)$ (notice that
$T^{2}/s_{0}(0) \ll 1$ in the temperature range under consideration).
Quantitatively, it would be interesting to compare our result with that of
the low temperature expansion, once this becomes available (to determine
the ${\cal O}(T^{2})$ correction one needs the two-loop, ${\cal O}(p^{6})$,
calculation  of the parameters entering $G(T)$ \cite{LL}).\\

Using $f_{\pi}(T)$ and $<\bar{q}q>_{T}$ as input, plus $s_{0}(T)$
obtained from Eq.(10), it is possible to use the two FESR above, 
Eqs.(12)-(13), to obtain two independent determinations of 
$\mu_{\pi}(T)$. The results of this calculation, shown in Fig.(3), are 
in good agreement with each other, as well as with results from other
methods \cite{GL}-\cite{Schenk}. This is rather important, as it provides
strong support for the validity of the QCD sum rule program at finite
temperature.\\ 

{\bf Acknowledgements} The work of (CAD) and (MSF) has been supported in
part by the FRD (South Africa), and that of (ML) by Fondecyt (Chile)
under grant No.1950797.\\

\section*{Figure Captions}
\begin{description}
\item[Figure 1:] The ratios $f_{\pi}^{2}(T)/f_{\pi}^{2}(0)$,
Solid curve (a); $<\bar{q}q>_{T}/<\bar{q}q>$, dotted curve (b), both from 
\cite{Bar1}; and $s_{0}(T)/s_{0}(0)$, dashed curve (c), from the FESR Eq.
(10).
\item[Figure 2:] The ratio of the GMOR relation, Eq.(14).
\item[Figure 3:] The ratio $\mu_{\pi}(T)/\mu_{\pi}(0)$ obtained by solving
the FESR Eq.(12), curve (a), and Eq.(13), curve (b). In both cases,
$f_{\pi}(T)$ and $<\bar{q}q>_{T}$ from \cite{Bar1} were used as input.
\end{description}
\end{document}